\documentclass[10pt,conference]{IEEEtran}
\usepackage{geometry} 
\usepackage[parfill]{parskip}    
\usepackage{graphicx}
\usepackage{amssymb}
\usepackage{epstopdf}
\newcommand{\emp}[1]{{\bf #1}}

\newtheorem{teiri}{\emp{Theorem}}[section]
\newtheorem{kei}{\emp{Corollary}}[section]

\newtheorem{hodai}{\emp{Lemma}}[section]
\newtheorem{teigi}{\emp{Definition}}[section]
\newtheorem{rei}{\emp{Example}}[section]
\newtheorem{chui}{\emp{Remark}}[section]

\newcommand{\spr}[1]{{\bf #1}}

\newcommand{\vep}{\varepsilon}
\newcommand{\vph}{\varphi}

\newcommand{\cC}{{\cal C}}

\newcommand{\cM}{{\cal M}}
\newcommand{\cN}{{\cal N}} 

\newcommand{\cR}{{\cal R}} 

\newcommand{\cX}{{\cal X}}
\newcommand{\cY}{{\cal Y}}

\newcommand{\ssx}{\spr{x}}

\newcommand{\nth}{\frac{1}{n}}

\newcommand{\bteiri}{\begin{teiri}}
\newcommand{\eteiri}{\end{teiri}}
\newcommand{\bkei}{\begin{kei}}
\newcommand{\ekei}{\end{kei}}
\newcommand{\brei}{\begin{rei}}
\newcommand{\erei}{\end{rei}}
\newcommand{\bhodai}{\begin{hodai}}
\newcommand{\ehodai}{\end{hodai}}
\newcommand{\bteigi}{\begin{teigi}}
\newcommand{\eteigi}{\end{teigi}}
\newcommand{\bchui}{\begin{chui}}
\newcommand{\echui}{\end{chui}}
\newcommand{\beq}{\begin{equation}}
\newcommand{\eeq}{\end{equation}}
\newcommand{\beqn}{\begin{eqnarray}}
\newcommand{\eeqn}{\end{eqnarray}}
\newcommand{\beqns}{\begin{eqnarray*}}
\newcommand{\eeqns}{\end{eqnarray*}}

\newcommand{\map}{\vph_n: \cX^n \to \cY^n}
\newcommand{\mapMtoY}{\vph_n: \cM_{M_n} \to \cY^n}
\newcommand{\mapXtoM}{\vph_n: \cX^n \to \cM_{M_n}}

\title{Polymatroids with Network Coding}
\begin{document}

\author{\authorblockN{Te Sun Han}
\authorblockA{Waseda University\\
Tokyo, Japan\\
Email: han@aoni.waseda.jp}}
%

\maketitle

\begin{abstract}
The problem of network coding for multicasting  a single source to multiple  sinks has first been studied by Ahlswede, Cai, Li and Yeung in 2000, in which they have established the celebrated max-flow mini-cut theorem on non-physical information flow over a network of independent channels. 
On the other hand, in 1980, Han has studied the case with correlated multiple  sources and a single sink from the viewpoint of polymatroidal functions in which a necessary and sufficient condition has been demonstrated for reliable transmission over the network. This  paper presents an attempt
to unify both cases, which leads to establish a necessary and sufficient condition for reliable 
transmission over a noisy network for multicasting  all the correlated  multiple  sources to all the multiple  sinks. Furthermore, we address also the problem of transmitting ``independent" sources over a multiple-access-type of network as well as over a broadcast-type of network, which reveals that the (co-) polymatroidal structures are intrinsically involved in these types of network coding.

\end{abstract}

\section{Introduction}\label{intro-geri1}
On the other hand, in 1980, Han \cite{han-cover} had studied the case with {\em correlated} multiple  sources and a single sink from the viewpoint of polymatroidal functions in which a necessary and sufficient condition has been demonstrated for reliable transmission over a  network. 

This  paper presents an attempt
to unify both cases and to generalize it to  quite a general case with stationary ergodic correlated sources and {\em noisy} channels (with {\em arbitrary} nonnegative real values of capacity that are {\em not} necessarily {\em integers}) satisfying the strong converse property (cf. Verd\'u and Han \cite{verdu-han}, Han \cite{han-book}),
which leads to establish a necessary and sufficient condition for reliable 
transmission over a  {\em noisy} network for multicasting  correlated  multiple  sources altogether to every multiple  sinks. 

It should be noted here that 
in such a   situation with {\em correlated} multiple  sources, the central issue turns out to be how to construct the 
{\em matching condition} between source and channel (i.e., joint source-channel coding), instead of  of  the traditional concept of {\em 
capacity region} (i.e., channel coding), although in the special case with {\em non-correlated} independent multiple sources the problem reduces again to how to describe the capacity region.

Several network models with correlated multiple  sources have been studied by some people, e.g.,
by Barros and Servetto \cite{bar-serv}, Ho, M\'edard, Effros and Koetter \cite{koetter}, 
Ho, M\'edard,  Koetter, Karger, Effros, Shi and Leong \cite{koetter-1}, 
Ramamoorthy, Jain, Chou and Effros \cite{effros}.
Among others, \cite{koetter}, \cite{koetter-1} and \cite{effros} consider (without attention to the {\em converse} part)  {\em error-free} network coding 
for two (or possibly {\em multiple}) stationary memoryless  correlated sources with a single 
(or possibly {\em multiple}) sink(s) to study
the error exponent problem,
where we notice that 
 %
  all the arguments in \cite{koetter}, \cite{koetter-1} and \cite{effros}
  can be validated  only  within  the  class of stationary memoryless sources of {\em  integer} bit rates and 
 {\em error-free} channels (i.e., the {\em identity} mappings) all with {\em one bit} capacity (or {\em integer bits} capacity,
 which is allowed by introducing {\em multiple} edges);  these restrictions are needed solely to invoke ``Menger's theorem" in graph theory. 
   The main result in the present paper is quite free from such seemingly severe restrictions, because we can dispense with the use of Menger's theorem.

On the other hand,  \cite{bar-serv} revisits the same model as in Han \cite{han-cover}, while
 \cite{effros} mainly focuses on  the network with two correlated sources and two sinks to discuss 
the separation problem of distributed source coding  and network coding,
where, in addition, cases of two-sources three-sinks, and three-sources two-sinks are also studied.
 It should be noted that, in the case of networks with correlated multiple  sources, such a {\em separation} problem is another central issue. 
In this paper, we describe  a  necessary and sufficient
condition for separability in the case with multiple sources and multiple sinks.
(cf. Remark \ref{chui:geriko1}). 

On the other hand, we may consider another network model with {\em independent} multiple  sources
but with multiple  sinks each of which is required to reliably reproduce a prescribed subset of the 
multiple  sources that depends on each sink. However, the problem with this general model looks quite hard, although, e.g., 
 Yan, Yeung and Zhang \cite{yan} and  Song, Yeung and Cai \cite{song-yeung} 
have demonstrated the entropy  characterizations of the capacity region, which  still contain 
 limiting operations and are not computable.
Incidentally, Yan, Yang and Zhang \cite{zhang} have considered, as a computable special case,  degree-2 three-layer networks with 
$K$-pairs transmission requirements to derive the explicit capacity region.
 In this paper, for the same reason, we  focus on the case in which all the correlated multiple sources 
is to be multicast to all the multiple sinks  and  derive a simple necessary and sufficient matching condition
in terms of conditional entropy rates and capacity functions. This case can be regarded as   the network counterpart of the non-network compound Slepian-Wolf system \cite{csis-kor}.

 We notice here the following; although throughout in the paper we are encountered with the subtleties coming from the general channel and source characteristics assumed, the main logical stream remains essentially  unchanged
 if  we consider  simpler models, e.g.,  such as stationary  correlated Markov sources
 together with stationary memoryless noisy channels. This means that considering only simple cases
 does not help so much at both of the conceptual and  notational levels of the arguments.
 For this reason, we preferred here the compact general settings.

The present paper consists of six sections. In Section \ref{ss:HT_LD} notations and preliminaries are described, and 
in Section \ref{ss:HT_LD_C} we state the main result as well as its proof.
In Section \ref{ss:geri-ex1} two examples are shown. Section \ref{alter-cond} provides another type of necessary and sufficient condition for transmissibility.
Also, some  detailed comments on the previous papers are given.
Finally, in Section \ref{saigo-per} we address the routing capacity problems with
polymatroids and co-polymatroids.
%

\section{Preliminaries and Notations \label{ss:HT_LD}}

\medskip

\noindent
\qquad{\em A. Communication networks} 

\medskip

Let us consider an acyclic directed graph $G=(V,E)$ where 
$V =\{1,2,\cdots, |V|\}$ $
(|V| <+\infty)$, $E\subset V \times V,$ but
$(i,i)\not\in E$ for all $i\in V$. Here, elements of $V$ are called {\em nodes}, and
elements $(i,j)$ of $E$  are called {\em edges} or {\em channels}
from $i$ to $j$. Each edge $ (i,j)$ is assigned the {\em capacity}  $c_{ij}\ge 0$,
which specifies the maximum amount of information flow passing through the
channel $(i, j)$. If we want to emphasize the graph thus capacitated, we write it as $G=(V, E,C)$ where $C=(c_{ij})_{(i,j)\in E}$. A graph $G=(V, E,C)$ is sometimes called a (communication) network, and indicated also by ${\cal N}=(V, E,C)$.
We consider two fixed subsets $\Phi, \Psi$ of $V$ such that 
$\Phi \cap \Psi = \emptyset$  (the empty set) with
\[
\Phi = \{s_1, s_2, \cdots, s_p\},
\]
\[
\Psi = \{t_1, t_2, \cdots, t_q\},
\]
where  elements of $\Phi$ are called  {\em source nodes}, while elements of $\Psi$ are called {\em sink nodes}.
Here, to avoid subtle irregularities, we assume that there are no edges $(i,s)$ such that $s \in \Phi.$

Informally, our problem is  how to simultaneously transmit the information generated at the source nodes in $\Phi$ altogether to all the sink nodes in $\Psi$.
More formally, this problem is described as in the following subsection.
%
%

\medskip

\noindent
\qquad{\em B. Sources and channels} 

\medskip

Each source node $s \in \Phi$  generates a stationary and  ergodic source process
\beq\label{eq:geri2}
X_s = (X_s^{(1)}, X_s^{(2)}, \cdots),
\eeq
where $X_s^{(i)} (i=1,2,\cdots)$  takes values in finite source alphabet $\cX_s$.
Throughout in this paper we consider the case in which the whole joint process $X_{\Phi} \equiv (X_s)_{s \in \Phi}$ 
is  stationary and ergodic.
It is then evident that the joint process $X_{T} \equiv (X_s)_{s \in T}$ is also stationary and ergodic for 
any $T$ such that $\emptyset \neq T \subset \Phi $.
The component processes $X_s\  (s\in \Phi) $  may be correlated.
We write $X_T$ as 
\beq\label{eq:geri3}
X_T = (X_T^{(1)}, X_T^{(2)}, \cdots)
\eeq
and put
\beq\label{eq:geri4}
X_T^n = (X_T^{(1)}, X_T^{(2)}, \cdots, X_T^{(n)}),
\eeq
where $X_T^{(i)}\  (i=1,2,\cdots)$ takes values in $ \cX_T\equiv \prod_{s\in T}\cX_s$.



On the other hand, it is assumed that all the channels $(i,j)\in E,$ specified by the transition probabilities 
%
$w_{ij}: A_{ij}^n \to B_{ij}^n$ with finite input alphabet $A_{ij}$ and
finite output alphabet $B_{ij}$, 
 are {\em statistically  independent} and satisfy the {\em strong converse } property (see Verd\'u and Han \cite{verdu-han}). It should be noted here that stationaty and memoryless ({\em noisy} or {\em noiseless}) channels with finite
 input/output alphabets satisfy, as  very special cases,  this property (cf. Gallager \cite{gall}, Han \cite{han-book}).
 Barros and Servetto \cite{bar-serv} have considered the case of stationary and  memoryless sources/channels with finite alphabets.
 The following lemma plays a crucial role in establishing the relevant converse of the main result:
 
 \bhodai\label{hoda-strong-cap}{\rm (Verd\'u and Han \cite{verdu-han})}
{\rm 
The channel capacity $c_{ij}$ of a channel $w_{ij}$ satisfying the strong converse property with finite input/output alphabets is given by
\[
c_{ij} = \lim_{n\to\infty} \nth \max_{X^n}I(X^n;Y^n),
\]
where $X^n, Y^n$ are the input and the output of the channel $w_{ij}$, respectively, and 
$I(X^n; Y^n)$ is the mutual information (cf. Cover and Thomas \cite{cover-thomas}).
\QED}
\ehodai

\medskip

\noindent
\qquad{\em C. Encoding and decoding} 

\medskip

In this section let us state the necessary operation of encoding and decoding for network coding with correlated multiple sources to be multicast to multiple sinks. 

With  arbitrarily
small $\delta>0$ and   $\vep>0$, we introduce an 
$(n, (R_{ij})_{(i,j)\in E},$ $
\delta, \vep )
$ code as the one as specified by (\ref{newji-1}) $\sim$ (\ref{newji-7}) below, where we use the notation $[1, M]$ to indicate $
\{1,2,\cdots, M\}$. How to construct a ``good" $(n, (R_{ij})_{(i,j)\in E},$ $
\delta, \vep )$ code will be shown in Direct part of the proof of Theorem \ref{teiri:newji-1}.

\medskip

\noindent
\quad{\em 1)} For all $(s,  j)\ (s \in \Phi)$, the encoding function is
\beq\label{newji-1}
f_{sj}: \cX^n_s\to [1, 2^{n(R_{sj}-\delta)}],
\eeq
where the output of $f_{sj}$ is carried over to the encoder $\varphi_{sj}$ of
channel $w_{sj}$, while the decoder $\psi_{sj}$ of $w_{sj}$ outputs an estimate of the output of $f_{sj}$, which is specified by the stochastic composite function: 
\beq\label{newji-2}
h_{sj}\equiv \psi_{sj}\circ w_{sj}\circ \varphi_{sj}\circ f_{sj}: \cX^n_s\to [1, 2^{n(R_{sj}-\delta)}];
\eeq
\medskip
\noindent
\quad{\em 2)} For all $(i,  j)$ $ ( i \not\in \Phi)$,
the encoding function is
\beq\label{newji-3}
f_{ij}: \prod_{k:(k,i)\in E} [1, 2^{n(R_{ki}-\delta)}]
\to [1, 2^{n(R_{ij}-\delta)}],
\eeq
where the output of $f_{ij}$ is carried over to the encoder $\varphi_{ij}$ of
channel $w_{ij}$, while the decoder $\psi_{ij}$ of $w_{ij}$ outputs an estimate of the output of $f_{ij}$, which is specified by the stochastic composite function: 
\beqn\label{newji-4}
\lefteqn{h_{ij}\equiv}\nonumber\\
 & \psi_{ij}\circ w_{ij}\circ \varphi_{ij}\circ f_{ij}: \prod_{k:(k,i)\in E} [1, 2^{n(R_{ki}-\delta)}]
\nonumber \\ &\qquad \qquad\qquad \to
[1, 2^{n(R_{ij}-\delta)}].
\eeqn
Here, if $\{k:(k,i)\in E\}$ is empty, we use the convention that $f_{ij}$ is an arbitrary constant function taking a value in  $[1, 2^{n(R_{ij}-\delta)}]$;
\medskip

\noindent
\quad{\em 3)} For all $t\in \Psi$, the decoding function is 
\beq\label{newji-5}
g_t: \prod_{k:(k,t)\in E} [1, 2^{n(R_{kt}-\delta)}]
\to \cX^n_{\Phi}.
\eeq
\noindent
\quad{\em 4) Error probability} 

\medskip
\noindent
All sink nodes $t\in \Psi$ are required to reproduce  a ``good" estimate  $\hat{X}^n_{\Phi,t}$
($\equiv$ the output of the decoder $g_t$) 
 of $X^n_{\Phi}$, through the network
 $\cN = (V,E,C)$, so that the error probability 
 $\Pr\{\hat{X}^n_{\Phi,t} \neq X^n_{\Phi}\}$ be as small as possible.
Formally, 
for all $t\in \Psi$,  the probability  $\lambda_{n,t}$ of decoding error committed at sink $t$
is required to satisfy
\beq\label{newji-7}
\lambda_{n,t} \equiv \Pr\{\hat{X}^n_{\Phi,t} \neq X^n_{\Phi}\} \le \vep
\eeq
for all sufficiently large $n$.
Clearly, $\hat{X}^n_{\Phi,t}$  are the random variables induced by 
$X^n_{\Phi}$ that were generated at all source nodes $s \in \Phi$.

\medskip

We now need the following definitions.

\bteigi [{\it rate achievability}]\label{teigi-newji-1}
{\rm
If there exists an $(n, (R_{ij})_{(i,j)\in E},$ $
\delta, \vep )
$ code  for any arbitrarily small $\vep>0$ as well as any sufficiently small $\delta>0$,
and for all sufficiently large $n$, then we say that the rate 
$(R_{ij})_{(i,j)\in E}$ is {\em achievable} for the network
$G=(V, E)$.
}
\QED
\eteigi
\bteigi [{\it transmissibility}]\label{teigi-newji-2}
{\rm
If, for any small $\tau>0$,  the augmented capacity rate $(R_{ij}=c_{ij}+ \tau)_{(i,j)\in E}$ is
achievable, then we say that the source $X_{\Phi}$ is {\em transmissible} over the network $\cN=(V,E, C),$ where $c_{ij} + \tau $ is called the $\tau$-capacity of channel $(i,j).$
}
\QED 
\eteigi
 The proof  of Theorem \ref{teiri:newji-1} (both of the converse part and the direct part) are based  on these definitions. 
%

\medskip

\noindent
\qquad{\em D. Capacity functions}  
\medskip

Let $\cN = (V,E,C)$ be a network. 
For any subset $M\subset V$ we say that $(M, \overline{M})$ 
(or simply, $M$) is a {\em cut} and 
$$E_M \equiv \{(i,j) \in E | i \in M, j\in \overline{M}\}$$
 the {\em cutset} of $(M, \overline{M})$
(or simply, of $M$), where $\overline{M}$ denotes the complement of $M$ in $V$.
Also, we call 
\beq\label{eq:newji-1}
c(M,\overline{M}) \equiv \sum_{ 
(i,j) \in E ,  i \in M,  j \in \overline{M}} c_{ij}
\eeq
the value of the cut $(M, \overline{M})$. Moreover, 
for any subset $S$ such that 
$\emptyset \neq S \subset \Phi$ (the source node set) and
for any $t\in \Psi$ (the sink node sets),
define
\beq\label{eq:newji-2}
\rho_t (S)= \min_{M: S \subset  M, t\in \overline{M}} c(M,\overline{M});
\eeq
\beq\label{eq:newji-34}
\rho_{\cN} (S)= \min_{t\in \Psi} \rho_t (S).
\eeq
We call this $\rho_{\cN} (S)$ the capacity function of
 $S\subset  V$ for the network $\cN = (V,E,C)$.
\bchui\label{chui-aho1}
{\rm
A set function $\sigma(S)$ on $\Phi$ is called a co-polymatroid
\footnote{In Zhang, Chen, Wicker and Berger \cite{xzhang}, 
the {\em co}-polymatroid here is called 
the
{\em contra}-polymatroid.} (function) if it holds that
\beqns
\mbox{1)} & & \sigma (\emptyset)  =  0,\\
\mbox{2)}& &\sigma (S) \le  \sigma ( T) \quad (S \subset T),\\
\mbox{3)} & & \sigma (S  \cap T) + \sigma (S \cup T) \ge  \sigma (S) + \sigma (T).
\eeqns
It is not difficult to check that $\sigma(S) = H(X_S|X_{\overline{S}})$  is a co-polymatroid
(see, Han \cite{han-cover}).
On the other hand, a set function $\rho (S)$ on $\Phi$ is called a polymatroid if it holds that
\beqns
\mbox{$1^{\prime}$)} & & \rho (\emptyset)  =  0,\\
\mbox{$2^{\prime}$)}& &\rho (S) \le  \rho ( T) \quad (S \subset T),\\
\mbox{$3^{\prime}$)} & & \rho (S  \cap T) + \rho (S \cup T) \le  \rho (S) + \rho (T).
\eeqns
It is also not difficult to check that 
for each $t\in \Psi$ the function $\rho_t (S)$ in (\ref{eq:newji-2})
is a polymatroid (cf. Han \cite{han-cover}, Meggido \cite{meggido}),
but $\rho_{\cN} (S)$ in (\ref{eq:newji-34})) is not necessarily a polymatroid.
These properties have been fully invoked in establishing the {\em matching condition} between source and channel for the special case of $|\Psi|=1$ ( cf. Han \cite{han-cover}).
\QED
}
\echui
With these preparations we will demonstrate the main result in the next section.
%
%
%
\section{Main Result \label{ss:HT_LD_C}}

The problem that we deal with here is not that of establishing the ``capacity region"
as usual, because the concept of ``capacity region" does not make sense for the general network with correlated sources. Instead,   we are interested in the {\em matching} problem between  the correlated source $X_{\Phi}$ and the network 
$\cN = (V,E,C)$ (transmissibility: cf. Definition \ref{teigi-newji-2}).
Under what condition is such a matching possible? This is the key problem here.
An answer to this question is just our main result to be stated here.
\bteiri\label{teiri:newji-1}
{\rm
The source $X_{\Phi}$ is transmissible over the network 
$\cN = (V,E,C)$ if and only if 
\beq\label{eq:newji-3}
H(X_S|X_{\overline{S}}) \le \rho_{\cN}(S)\quad (\emptyset \neq \forall S \subset \Phi)
\eeq
holds. \QED
}
\eteiri
%
\bchui
{\rm
The case of $|\Psi| =1$ was investigated by Han \cite{han-cover}, 
and subsequently revisited by Barros and Servetto \cite{bar-serv},
while the case of $|\Phi| =1$ was investigated by Ahlswede, Cai, Li and Yeung \cite{al-yeung}.
%
}
%
\QED
\echui

\bchui\label{chui-benpi}
{\rm
If the sources are {\em mutually independent}, (\ref{eq:newji-3}) reduces to
\[
\sum_{i \in S}H(X_i) \le 
\rho_{\cN}(S) \quad (\emptyset \neq \forall S \subset \Phi).
\]
Then, setting the rates as $R_i = H(X_i) $  we have another equivalent form:
\beq\label{eq:miyako-1}
\sum_{i \in S}R_i \le 
\rho_{\cN}(S) \quad (\emptyset \neq \forall S \subset \Phi).
\eeq
This specifies the {\em capacity region} of independent message rates in the traditional sense.
In other words, in case  the sources are  independent, the concept of capacity region makes sense.
In this case too, channel coding looks like for
{\em non-physical}  flows (as for the  case of $|\Phi| =1$, see Ahlswede, Cai,  Li and Yeung
\cite{al-yeung}; and as for the case of $|\Phi| >1$ see, e.g., Koetter and Med\'ard  
 \cite{koetter-med}, Li and Yeung \cite{li-yeung}). It should be noted  that formula (\ref{eq:miyako-1}) is {\em not} derivable by
a naive extension of the arguments as used in the case of single-source ($|\Phi|=1$), irrespective of  the comment in \cite{al-yeung}.
}
\QED
\echui

%

\noindent
{\em Proof of Theorem\ref{teiri:newji-1}:\  {\rm The proof is based on joint typicality, strong converse property of the channel, acyclicity of the network, ergodicity of the source, random coding arguments,  Fano's inequality, subtle classification of the error patterns, and so on.
As for the details, see Han \cite{han-shannon}).}
}
\bigskip



%
\section{Examples\label{ss:geri-ex1}}
In this section we show two examples of Theorem \ref{teiri:newji-1} with 
$\Phi =\{s_1. s_2\}$ and $\Psi =\{t_1. t_2\}$.

\medskip

\noindent

\noindent
{\em Example 1}.\ Consider the network as in Fig.1(called the {\em butterfly}) where all the solid edges have capacity 1
and the independent sources $X_1, X_2$ are binary and uniformly distributed. The capacity function of this network is computed as follows:
\beqns
\rho_{t_1}(\{s_2\}) & = & \rho_{t_2}(\{s_1\}) = 1,\\
\rho_{t_1}(\{s_1\}) & = & \rho_{t_2}(\{s_2\}) = 2,\\
\rho_{t_1}(\{s_1,s_2\}) & = & \rho_{t_2}(\{s_1,s_2\}) = 2;
\eeqns
\beqns
\rho_{{\cal N}}(\{s_1\}) & = & \min (\rho_{t_1}(\{s_1\}),\rho_{t_2}(\{s_1\}))=1,\\
\rho_{{\cal N}}(\{s_2\}) & = & \min (\rho_{t_1}(\{s_2\}),\rho_{t_2}(\{s_2\}))=1,\\
\rho_{{\cal N}}(\{s_1,s_2\}) & = & \min (\rho_{t_1}(\{s_1,s_2\}),\rho_{t_2}(\{s_1,s_2\}))\\
& & =2.
\eeqns
On the other hand, 
\beqns
H(X_1|X_2) & = & H(X_1) =1,\\
H(X_2|X_1) & = & H(X_2) =1,\\
H(X_1X_2) & = & H(X_1) + H(X_2) =2.
\eeqns

\begin{figure}[htbp]
   \centering
   \includegraphics[bb= 0 0 200 310, scale=0.4]{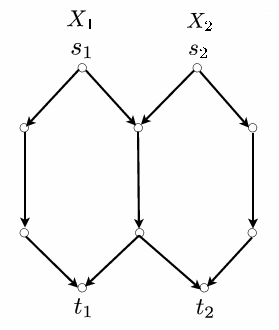} 
   \caption{Example 1}
\end{figure}

Therefore, condition (\ref{eq:newji-3}) in Theorem \ref{teiri:newji-1} is satisfied with equality, so that the sourse is transmissible over the network. Then, how to attain this transmissibility? That is depicted in Fig.2 where $\oplus$ denotes the exclusive OR. Fig. 3 depicts the corresponding  capacity region,
which is within the framework of the previous work (e.g., see Ahlswede {\em et al.} \cite{al-yeung}).

\begin{figure}[htbp]
   \centering
   \includegraphics[bb= 0 0 200 310, scale=0.4]{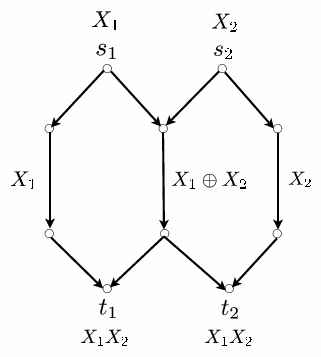} 
   \caption{Coding for Example 1}
\end{figure}

\begin{figure}[htbp]
   \centering
   \includegraphics[bb= 0 0 200 310, scale=0.4]{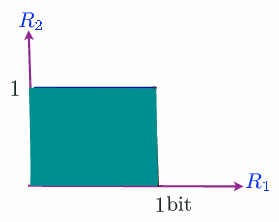} 
   \caption{Capacity region for Example 1}
\end{figure}

\medskip

\noindent
{\em Example 2}. \ Consider the network with {\em noisy} channels as in Fig.4 where the solid edges have capacity 1 and the broken edges have capacity $h(p)<1 $. Here, $h(p)$ ($0< p < \frac{1}{2}$) is the binary entropy defined by
$ h(p) = -p\log_2 p -(1-p)\log_2 (1-p).$ The source $(X_1, X_2)$ 
generated at the nodes $s_1, s_2$ is the binary symmetric source with crossover probability $p$,
i.e.,
\beqns
\Pr\{X_1=1\}& =& \Pr\{X_1=0\} =\Pr\{X_2=1\}\\
& =& \Pr\{X_2=0\}=\frac{1}{2},
\eeqns
\[
\Pr\{X_2=1|X_1=0\} =\Pr\{X_2=0|X_1=1\}=p.
\]
 Notice that 
$X_1, X_2$  are not independent.
The capacity function of this network is computed as follows:
\beqns
\rho_{t_1}(\{s_2\}) & = & \rho_{t_2}(\{s_1\}) = h(p),\\
\rho_{t_1}(\{s_1\}) & = & \rho_{t_2}(\{s_2\}) =1+h(p),\\
\rho_{t_1}(\{s_1,s_2\}) & = & \rho_{t_2}(\{s_1,s_2\}) = \min(1+2h(p), 2);
\eeqns
\beqns
\rho_{{\cal N}}(\{s_2\}) & = & \min (\rho_{t_1}(\{s_2\}),\rho_{t_2}(\{s_2\}))=h(p),\\
\rho_{{\cal N}}(\{s_1\}) & = & \min (\rho_{t_1}(\{s_1\}),\rho_{t_2}(\{s_1\}))=h(p),\\
\rho_{{\cal N}}(\{s_1,s_2\}) & = & \min (\rho_{t_1}(\{s_1,s_2\}),\rho_{t_2}(\{s_1,s_2\}))\\
& & =2.
\eeqns
On the other hand, 
\beqns
H(X_1|X_2) & = &h(p),\\
H(X_2|X_1) & = &h(p),\\
H(X_1X_2) & = & 1+h(p).
\eeqns

\begin{figure}[htbp]
   \centering
   \includegraphics[bb= 0 0 200 310, scale=0.4]{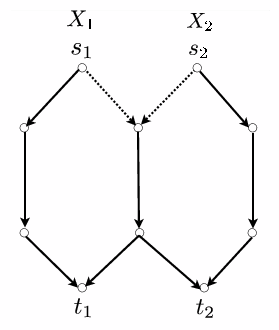} 
   \caption{ Example 2}
\end{figure}

\begin{figure}[htbp]
   \centering
   \includegraphics[bb= 0 0 200 350, scale=0.4]{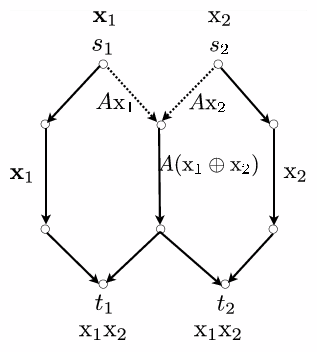} 
   \caption{Coding for  Example 2}
\end{figure}

Therefore, condition (\ref{eq:newji-3}) in Theorem \ref{teiri:newji-1} is satisfied with strict inequality, so that the source is transmissible over the network. Then, how to attain this transmissibility? That is depicted in Fig.5 where  $\ssx_1, \ssx_2$ are $n$ independent copies of $X_1, X_2$, respectively,
and $A$ is an $m\times n$ matrix ($m=nh(p)<n)$.
Notice that the entropy  of $\ssx_1\oplus \ssx_2$ (componentwise exclusive OR) is  $nh(p)$ bits and hence it is possible to recover $\ssx_1\oplus \ssx_2$ from $A(\ssx_1\oplus \ssx_2)$ (of length $m =nh(p)$) with asymtoticaly negligible probability of decoding error,  provided that $A$ is appropriately chosen 
(see K\"orner and Marton \cite{korn-mart}). It should be remarked that this example cannot be justified by the previous works such as Ho {\em et al.} \cite{koetter}, Ho {\em et al.} \cite{koetter-1}, and
Ramamoorthy {\em et al.} \cite{effros}, because all of them assume  {\em noiseless} channels with 
capacity of  {\em one bit}, i.e.,
this example is outside the previous framework.

\section{Alternative Transmissibility Condition}\label{alter-cond}
In this section we demonstrate an alternative transmissibility condition equivalent to the 
necessary and sufficient condition  (\ref{eq:newji-3}) given 
in Theorem \ref{teiri:newji-1}. 

To do so, for each $t\in \Psi $ we define the polyhedron $\cC_t$ as the set of 
all nonnegative rates
$(R_s; s \in \Phi)$ such that 
\beq\label{eq:efros-san}
\sum_{i \in S}R_i \le 
\rho_{t}(S) \quad (\emptyset \neq \forall S \subset \Phi),
\eeq
where $\rho_{t}(S) $ is the capacity function as defined in (\ref{eq:newji-2}) of Section \ref{ss:HT_LD}.
Moreover, define the polyhedron $\cR_{\mbox{{\scriptsize SW}}}$ as the set of all nonnegative
rates
$(R_s; s \in \Phi)$ such that 
\beq\label{eq:efros-san-2}
H(X_S|X_{\overline{S}}) \le \sum_{i \in S}R_i  \quad (\emptyset \neq \forall S \subset \Phi),
\eeq
where $H(X_S|X_{\overline{S}})$ is the conditional entropy rate as defined in Section \ref{ss:HT_LD}.
Then, we have the following theorem on the transmissibility over the network $\cN=(V, E, C)$.
\bteiri\label{teiri:odoroi-te}
{\rm The following two statements are equivalent:
\beqn
 1)& & H(X_S|X_{\overline{S}}) \le \rho_{\cN}(S)\quad (\emptyset \neq \forall S \subset \Phi),
\nonumber\\ & & \label{eq:teruo-1}\\
 2) & & \cR_{\mbox{{\scriptsize SW}}}\cap \cC_t \neq \emptyset
\quad (\forall t\in \Psi).\label{eq:teruo-2}
\eeqn
}
\eteiri  
In order to prove Theorem \ref{teiri:odoroi-te} we need the following lemma:
\bhodai [{\rm Han \cite{han-cover}}]\label{hodai:kopit-1} 
{\rm 
Let $\sigma(S)$, $\rho(S)$ be a co-polymatroid and a polymatroid, respectively, 
as defined in Remark \ref{chui-aho1}. Then,  the necessary and sufficient condition for the 
existence of some nonnegative rates  $(R_s; s \in \Phi)$ such that
 \beq\label{eq:mannax-3}
 \sigma (S) \le \sum_{i \in S} R_i   \le   \rho (S)   \quad (\emptyset \neq \forall S \subset \Phi)
\eeq
is that
  \beq\label{eq:mannax-4}
\sigma (S) \le  \rho (S)\quad (\emptyset \neq \forall S \subset \Phi).
\eeq
}
\ehodai

\medskip

\noindent
{\em  Proof of Theorem \ref{teiri:odoroi-te}}:\ It is easy to see this.

\bchui\label{chui:toutou-2}
{\rm  
The necessary and sufficient condition of the form (\ref{eq:teruo-2}) appears ({\em without} the proof) in Ramamoorthy, Jain, Chou
 and Effros \cite{effros}, which they call the {\em feasibility}. They attribute the sufficiency part simply to Ho, M\'edard, Effros and Koetter \cite{koetter} 
 with $|\Phi| =2, |\Psi|=1$ (also, cf. 
Ho, M\'edard,  Koetter, Karger, Effros, Shi, and Leong \cite{koetter-1} with $|\Phi| =2, |\Psi|=1$),
 while attributing  the necessity part to Han \cite{han-cover}, Barros and  Servetto \cite{bar-serv}.
 However,  notice that 
 %
  all the arguments in \cite{koetter}, \cite{koetter-1} 
  (\cite{koetter} is included in \cite{koetter-1})
  can be validated  only  within  the class of stationary memoryless sources of {\em  integer} bit rates and 
 {\em error-free} channels (i.e., the {\em identity} mappings) all with {\em one bit} capacity
   (this restriction is needed to invoke ``Menger's theorem" in graph theory); 
 while the present  paper, without such ``seemingly" severe restrictions,  treats  ``general" acyclic networks, allowing  for general correlated stationary ergodic sources as well as general statistically  independent channels with each satisfying the strong converse property (cf. Lemma \ref{hoda-strong-cap}). Moreover, as long as we are concerned also  with 
 {\em noisy} channels, the way of approaching the problem as in \cite{koetter}, \cite{koetter-1} 
does {\em not} work as well, because in this noisy  case we have to cope with two kinds of error probabilities, 
 one due to error probabilities for source coding and the other due to error probabilities for network coding (i.e., channel coding); thus in the noisy channel case or in the noiseless channel case with {\em non-integer} capacities
 and/or i.i.d. sources of  {\em non-integer} bit rates, \cite{effros} cannot attribute the sufficiency part 
 of (\ref{eq:teruo-2}) to \cite{koetter}, \cite{koetter-1}. 
 
 It should be noted here also that   \cite{koetter} and \cite{koetter-1}, though demonstrating relevant error exponents (the {\em direct} part),  do not have the {\em converse} part.
 \QED
}
\echui 
\bchui [{\it Separation}]\label{chui:geriko1}
{\rm
Here, the term of {\em separation} is used to mean ``two step" separation of distributed source coding
({\em Step 1}: based on Slepian-Wolf theorem)
 and network coding with {\em independent} sources
 ({\em Step 2}: based on Theorem \ref{teiri:newji-1} with Remark \ref{chui-benpi}). 
%
%
%
Now suppose that 
 that there exist some nonnegative rates $R_i$ $ (i\in \Phi)$ such that
\beq\label{look-like-1}
H(X_S|X_{\overline{S}}) \le \sum_{i \in S}R_i \le 
\rho_{\cN}(S) \quad (\emptyset \neq \forall S \subset \Phi).
\eeq	
Then, the first inequality ensures {\em reliable} distributed source coding by virtue of  the 
theorem of Slepian and Wolf ({\em Step 1}), while the second
inequality ensures {\em reliable} network  coding by virtue of Thjeorem \ref{teiri:newji-1} with  Remark \ref{chui-benpi} for  independent sources ({\em Step 2}), which looks like for {\em non-physical}  flows,   with {\em independent} distributed sources of rates $R_i$ ($i \in \Phi$; see Remark \ref{chui-benpi}). 
Thus, (\ref{look-like-1}) is {\em sufficient} for separability.
 Condition (\ref{look-like-1}) is equivalently written as  
 \beq\label{eq:equiq1}
 \displaystyle{
 \cR_{\mbox{{\scriptsize SW}}}\cap 
\left(\bigcap_{t\in \Psi} \cC_t\right) \neq \emptyset
}
 \eeq
 for any general network $\cN$. Moreover, in view of Remark \ref{chui-benpi}, it  is not difficult to check that
 (\ref{eq:equiq1}) is also necessary.
 Thus, 
our conclusion is  that, in general,  condition (\ref{eq:equiq1}) is not only {\em sufficient} but 
also {\em necessary} for separability in the sense above stated.%
}
\QED
\echui

\bchui
{\rm
In this connection, \cite{effros} claims
 that, in the case with
$|\Phi | = |\Psi | = 2$ and with {\em rational} capacities as well as sources of {\em integer}
 bit rates, ``{\em separation}"  always holds, while
 in the case of $|\Phi|>2$ or $|\Psi|>2$ no
 conclusive claim is still not made. Although this seems to contradict Remark \ref{chui:geriko1}, 
 it should be merely due to the different definitions for separability.
 }
\echui

\bchui\label{chui:toutou-A3-1}
{\rm 
It is possible also to consider  network coding with {\em cost}. In this regard the reader may refer to, e.g., 
Han \cite{han-cover}, Ramamoorthy \cite{rama}, Lee {\em et al.} \cite{lee}.
\QED
}
\echui
\bchui\label{chui:toutou-A3}
{\rm 
So far we have focused on  the  case where the channels of a network are quite general but  are statistically
independent. On the other hand, we may think of the case where the channels are not necessarily statistically 
independent. This problem is quite hard in general. A typical tractable example of such networks would be a class of acyclic deterministic relay networks with no interference (called the Aref network)
in which the concept of ``channel capacity" is irrelevant. In this connection,  Ratnakar and
 Kramer  \cite{rat-kramer} have studied Aref networks with a single source and multiple  sinks, while 
Korada and Vasudevan \cite{korada-vasu} have studied Aref networks with multiple correlated sources and multiple sinks. The network capacity   formula as well as the network matching  formula obtained by them are in nice correspondence with the formula 
obtained by Ahlswede {\em et al.} \cite{al-yeung} as well as Theorem \ref{teiri:newji-1}
established in this paper, respectively.
}
\QED
\echui
%
 %
%
\section{Routing Capacity Regions}\label{saigo-per} 
%
So far we have considered the problem of multicasting multiple sources to multiple sinks
over a {\em noisy} network
in which all the sources may be  mutually {\em  correlated}. The fundamental toosl for this kind of 
reliable transmission are mainly {\em routing } and {\em coding} at each node of the network.
Along this line we have established Theorem \ref{teiri:newji-1} to give a necessary and sufficient condition for reliable transmission. On the other hand, several class of network coding may not need the operation of
{\em coding} but only that of {\em routing}. However, Theorem \ref{teiri:newji-1} does not provide any
explicit suggestions or answers in this respect. 

In this section, we address this problem with mutually {\em independent} sources specified only by their {\em rates} $R_i$'s. The set of all such achievable rates will be called the {\em routing capacity} of the network.
In what follows, as illustrative cases, we take three types of network routings, i.e.,
multiple-access-type of routing, broadcast-type of routing, and interference-type of routing.
In doing so, for each edge $(i,j) \in E$ we impose not only {\em upper} capacity $c_{ij}$ restriction but also {\em lower} capacity $ d_{ij}$ restriction ($0 \le d_{ij} \le c_{ij}$), which means that 
information  flows $g_{ij}$ passing
the channel $(i,j) \in E$ are restricted so that $d_{ij} \le g_{ij} \le c_{ij}$ for all  $(i,j)\in E$.
A motivation for the introduction of such lower capacities $ d_{ij}$ is that in some situations 
``informational outage" over a network  is to be avoided, for example.

\medskip

Let us first  state ``Hoffman's theorem" in a graph theory which is needed to discuss the routing capacities. 
For simplicity we put $D=(d_{ij})_{(i,j)\in E}$ like $C=(c_{ij})_{(i,j)\in E}$,
and also write as $[d_{ij}, c_{ij}]$.
\bteigi\label{teigi:rout1}
{\rm
Given a graph $G=(V,E, C, D)$, a flow $g_{ij}$ is said to be {\em circular} if 
 $d_{ij} \le g_{ij} \le c_{ij}$ for all  $(i,j)\in E$, and $ \sum_{j \in V}g_{ij}$ $= \sum_{j \in V}g_{ji}=0$
 for all $i \in V$ (the conservation law). \QED
 }
 \eteigi
 \bteiri\label{teiri:rout1}[Hoffman's theorem]
 {\rm
There exists  a circular flow $(g_{ij}: (i,j)\in E)$  if and only if
\beq\label{eq:rour-g1}
c(M,\overline{M}) \ge d(\overline{M}, M)\quad \mbox{for all subset}\  M\subset V,
\eeq
where  $c(M,\overline{M})$ was specified in (\ref{eq:newji-1}) and $d(\overline{M}, M)$
 is similarly defined by
replacing $c, M, \overline{M}$ by $d,  \overline{M}, M$, respectively, where $\overline{M}$ denotes  the complement of $M$ in $V$.\QED
}
\eteiri

\medskip

\noindent
\ {\em E. Multiple-access-type of  network routing}

\medskip

\noindent
Suppose that we are given a network $\cN =(V,E,C,D)$ as in Fig.6:

\begin{figure}[htbp]
   \centering
   \includegraphics[bb= 0 0 450 180, scale=0.3]{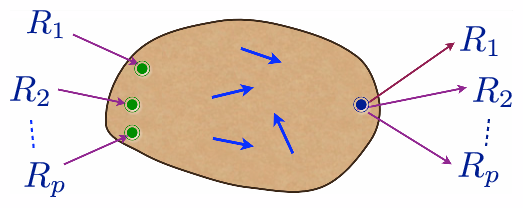} 
   \caption{Multiple-access-type of network}
\end{figure}

We modify this network by adding $p$ fictitious edges as in Fig.7:

\begin{figure}[htbp]
   \centering
   \includegraphics[bb= 0 0 300 280, scale=0.3]{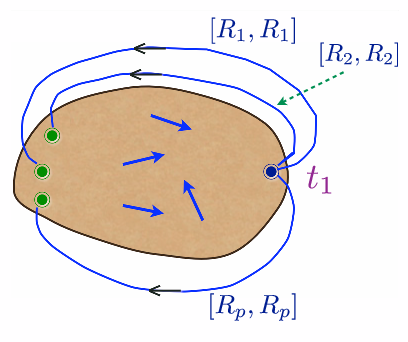} 
   \caption{Modified network $\cN^*$}
\end{figure}
 to obtain the modified network $\cN^* =$ $ (V, E^*, C, D,$ $ [R_1,R_1],\cdots, [R_p,R_p])$.
 We notice here that a rate $(R_1,\cdots, R_p)$ is achievable over the original network $\cN$ if and only if 
 the network $\cN$ has a circular flow.
 Thus, by writing down all the conditions in Hoffman's theorem for $\cN^*$, we obtain the following two kinds of inequalities:
 
 \begin{figure}[htbp]
   \centering
   \includegraphics[bb= 0 0 900 550, scale=0.25]{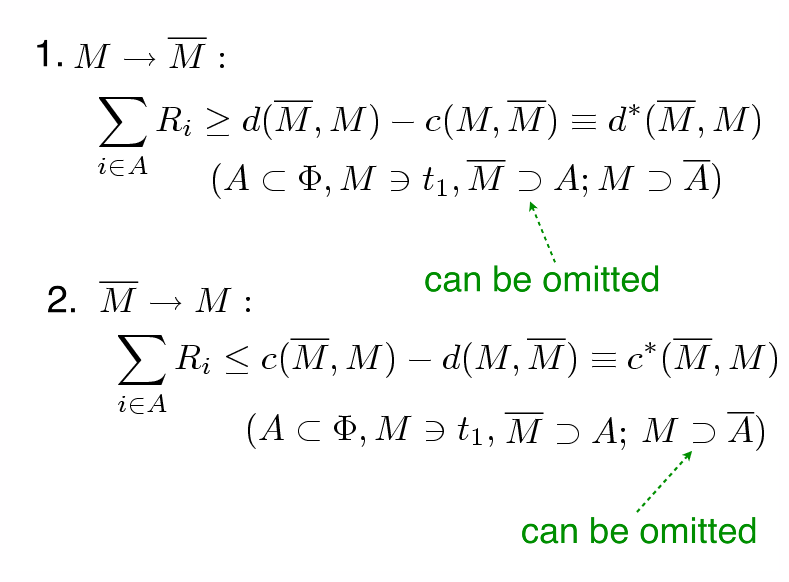} 
   \caption{Conditions for achievability}
\end{figure}

%
  
In order to express these conditions in a compact form, define the ``capacity function" as follows:
\bteigi\label{la:rot-13} \mbox{} For each subset $A$ of $\Phi$, let
  \beqn
  \rho_{\mbox{m}} (A) & \equiv & \min\{c^*(\overline{M},M)|M\ni t_1, \overline{M}\supset A\},\label{er:fogr1}
  \nonumber\\
  \sigma_{\mbox{m}} (A) & \equiv & \max\{d^*(\overline{M},M)|M\ni t_1, M\supset \overline{A}\}.
\label{er:fogr2}\nonumber
  \eeqn 
\eteigi
\bteiri\label{er:lit2}\mbox{} 
$\rho_{\mbox{m}} (A)$, $\sigma_{\mbox{m}} (A)$ are a polymatroid and a co-polymatroid, respectively.
\eteiri
%
\bteiri\label{er:lit3}\mbox{} There exists a circular flow if and only if 
\beq\label{re:poty1}
\sigma_{\mbox{m}} (A) \le \rho_{\mbox{m}} (A)\mbox{\quad for all }  A \subset \Phi, 
\eeq
and the routing capacity region of multiple-access-type of network is given by 
\beq\label{plo-1}
\sigma_{\mbox{m}} (A) \le \sum_{i\in A}R_i\le \rho_{\mbox{m}} (A)\mbox{\ for all }  A \subset \Phi.
\eeq
\eteiri
{\em Proof:} It suffices only to use Lemma \ref{hodai:kopit-1}. \QED
\bchui\label{chui:mnsat1}
{\rm 
It is easy to see that  $\sigma_{\mbox{m}}(A) =0$ for all $A \subset \Phi$ if $d_{ij}=0$ (for all $(i,j)\in E)$, and hence Theorem \ref{er:lit3} turns out  to be  a special case of Thorem \ref{teiri:newji-1} with independent sources. This implies
that in this case the capacity region given by Thorem \ref{teiri:newji-1} can actually be attained  only with {\em network  routing} but without {\em network coding}.
}
\echui

\noindent
\ {\em F. Broadcast-type of  network routing}


\noindent
Suppose that we are given a network $\cN =(V,E,C,D)$ as in Fig.9:

\begin{figure}[htbp]
   \centering
   \includegraphics[bb= 0 0 450 180, scale=0.3]{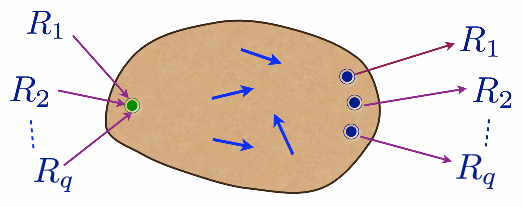} 
   \caption{Broadcast-type of network}
\end{figure}

We modify this network by adding $q$ fictitious edges as in Fig.10:

\begin{figure}[htbp]
   \centering
   \includegraphics[bb= 0 0 300 300, scale=0.3]{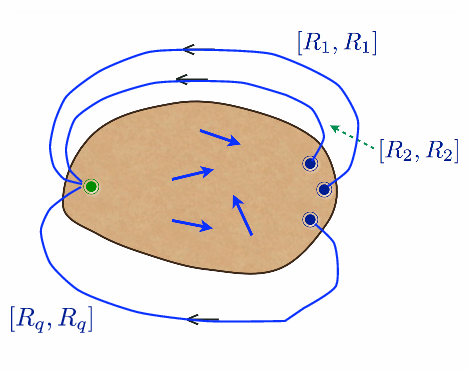} 
   \caption{Modified network $\cN^*$}
\end{figure}

 to obtain the modified network $\cN^* =$ $ (V, E^*, C, D,$ $ [R_1,R_1],\cdots, [R_q,R_q])$.
 Thus, in  a similar manner, we have the achievability conditions:

\begin{figure}[htbp]
   \centering
   \includegraphics[bb= 0 0 900 600, scale=0.25]{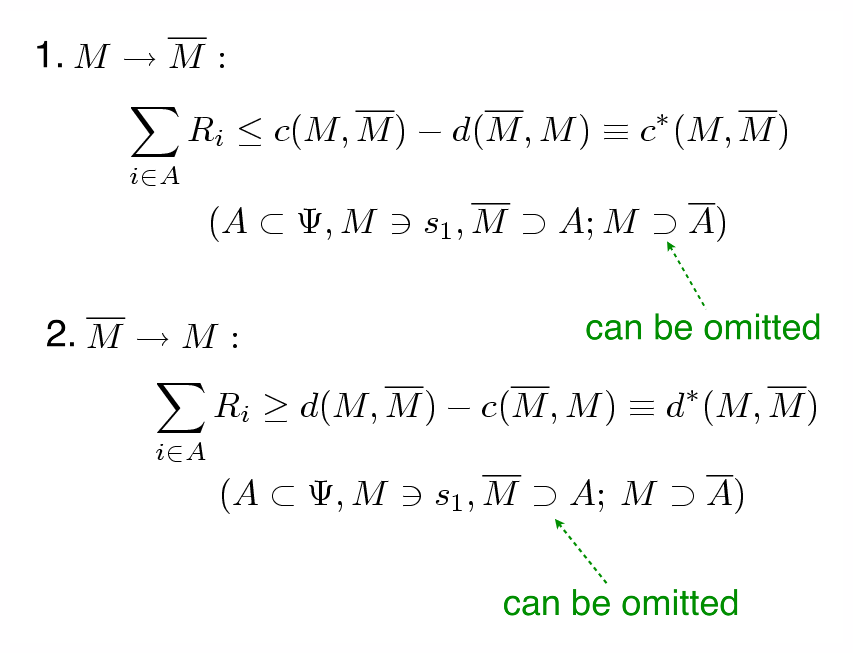} 
   \caption{Conditions for achievability}
\end{figure}
Accordingly,  the ``capacity function"  is defined as: 
\bteigi\label{la:rot-131} \mbox{} For each subset $A$ of $\Psi$, let
  \beqn
  \rho_{\mbox{b}} (A) & \equiv & \min\{c^*(M,\overline{M})|M\ni s_1, \overline{M}\supset A\},\label{er:fogr1}
  \nonumber\\
  \sigma_{\mbox{b}} (A) & \equiv & \max\{d^*(M,\overline{M})|M\ni s_1, M\supset \overline{A}\}.
\label{er:fogr2}\nonumber
  \eeqn 
\eteigi
\bteiri\label{er:lit212}\mbox{} 
$\rho_{\mbox{b}} (A)$, $\sigma_{\mbox{b}} (A)$ are a polymatroid and a co-polymatroid, respectively.
\eteiri
%
\bteiri\label{er:lit3kl}\mbox{} There exists a circular flow if and only if 
\beq\label{re:poty111}
\sigma_{\mbox{b}} (A) \le \rho_{\mbox{b}} (A)\mbox{\quad for all }  A \subset \Psi, 
\eeq
and the routing capacity region of broadcast-type of network is given by
\beq\label{plo-1yt}
\sigma_{\mbox{b}} (A) \le \sum_{i\in A}R_i\le \rho_{\mbox{b}} (A)\mbox{\ for all }  A \subset \Psi.
\eeq
\eteiri
{\em Proof:} It suffices only to use Lemma \ref{hodai:kopit-1}. \QED
\bchui\label{eq:kantep0}
{\em
The capacity region (\ref{plo-1yt}) remains unchanged if network coding is allowed in addition to network routing.\QED
}
\echui
 \noindent
\ {\em G. Interference-type of  network routing}


\noindent
Suppose that we are given a network $\cN =(V,E,C,D)$ as in Fig.12:

\begin{figure}[htbp]
   \centering
   \includegraphics[bb= 0 0 450 180, scale=0.3]{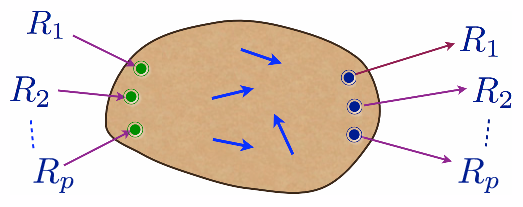}
   \caption{Interference-type of network}
\end{figure}

We modify this network by adding $p$ fictitious edges as in Fig.13 below.

\begin{figure}[htbp]
   \centering
   \includegraphics[bb= 0 0 450 320, scale=0.3]{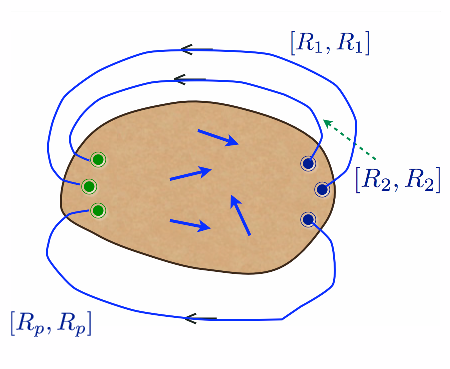} 
   \caption{Modified network $\cN^*$}
\end{figure}
 
In this case too, again by virtue of Hoffman's theorem, the achievability condition is summarized as
\beqn\label{rtmokp}
\max_{(A,B)} d^*(\overline{M},M) & \le & \sum_{i\in A_2}R_i - \sum_{i\in A_3}R_i \nonumber\\
&\le & \min_{(A,B)}c^*(M,\overline{M}),
\eeqn
where $\max_{(A,B)}$  (resp. $\min_{(A,B)}$)  denotes $\max$ (resp. $\min$) over all possible 
 cuts $(M, \overline{M})$ and  partitions  (with  $A_2, A_3$ fixed) 
 as was shown in Fig.14:

\begin{figure}[htbp]
   \centering
   \includegraphics[bb= 0 0 900 600, scale=0.25]{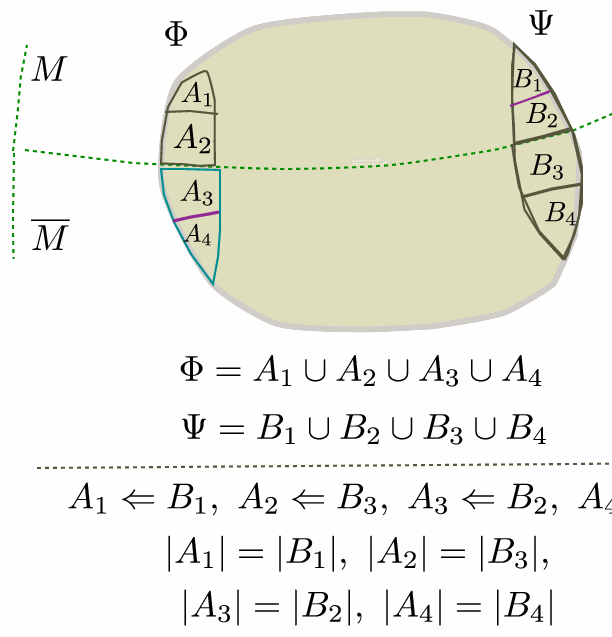} 
   \caption{Conditions for achievability}
\end{figure}

For simplicity, let us now consider the case  with $|\Phi | = | \Psi | =2$:

\begin{figure}[htbp]
   \centering
   \includegraphics[bb= 0 0 450 320, scale=0.3]{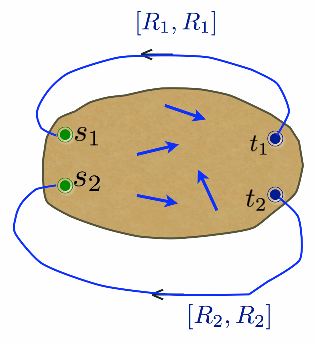} 
   \caption{Modified network $\cN^*$}
\end{figure}
Then, the achivability condition reduces to that as in Fig.16 below, where, 
on the contrary to the case of multiple-access-type of network routing
as well as to the case of broadcast-type of network routing,
the term $R_1 - R_2$ appears in addition to the term $R_1 + R_2$,
which suggests that the interference-type of network routing 
may be more complicated with  a rather pathological  behavior.

\begin{figure}[htbp]
   \centering
   \includegraphics[bb= 0 0 900 800, scale=0.25]{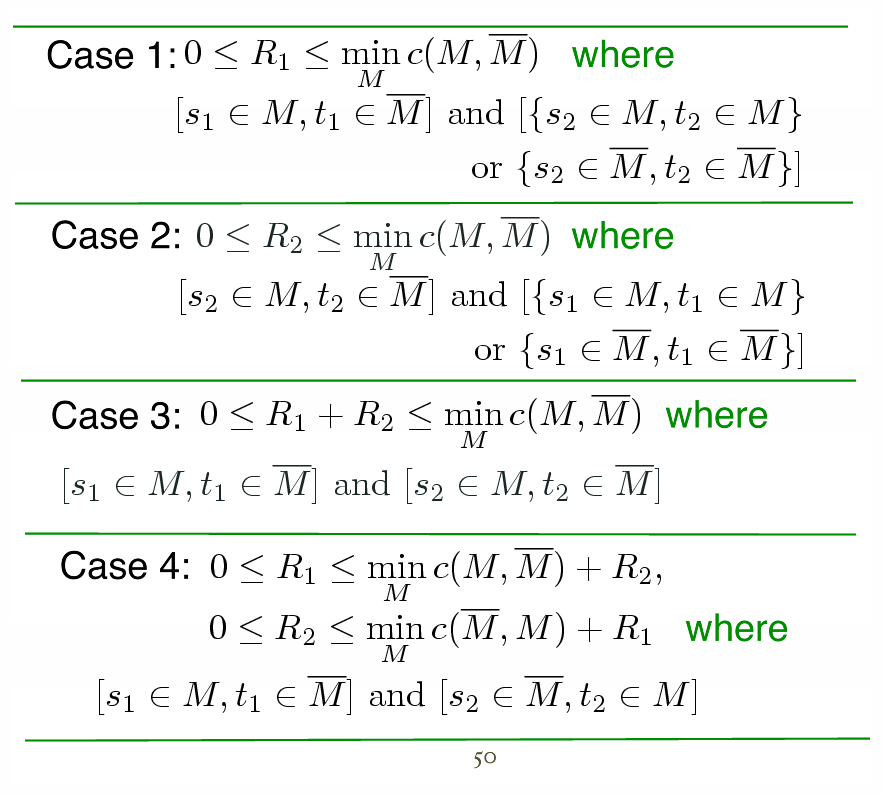} 
   \caption{Conditions for achievability with $d_{ij}=0$}
\end{figure}
The following Fig.17 shows a typical  region of Fig.16,
which looks like a rather untraditional shape.

\begin{figure}[htbp]
   \centering
   \includegraphics[bb= 0 0 450 420, scale=0.3]{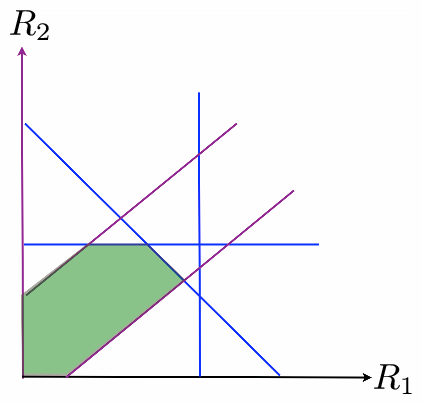} 
   \caption{``achievable" region}
\end{figure}

For example, let us consider the network $\cN$ as in Fig.18.
It is easy to check that the achievability condition in Fig. 16 for this network
reduces to $0\le R_1\le R_2\le 1.$
Fig.19 shows 
 two elementary directed cycles: one with $R_1=0, R_2=1$; the other with 
 $R_1=1, R_2=1$. The first flow actually specifies  a flow in $\cN$, 
 while the latter flow contains two fictitious edges and 
 hence does not specify any flow in $\cN$.  Thus, in the case of the interference-type of network,
 we have to take another approach to establish the routing capacity region.
An approach, though brute,  would be to list up all the elementary paths in $\cN$
connecting source $s_i$ and sink $t_i$ for every $i=1,2,\cdots, p$;
and, for every edge $e$ of ${\cal N}$, to check whether the total flow passing through  $e$ 
satisfies the upper and lower capacity constraints.
Incidentally, it should also be remarked   that any 
polymatroidal structure does not appear here. Anyway, Hoffman's theorem does not work 
as well for interference-type of networks. Thus, the routing capacity problem for 
interference-type of networks would be rather intractable and elusive, 
compared to multiple-access-type of networks and broadcast-type of networks.

\begin{figure}[htbp]
   \centering
   \includegraphics[bb= 0 0 450 480, scale=0.3]{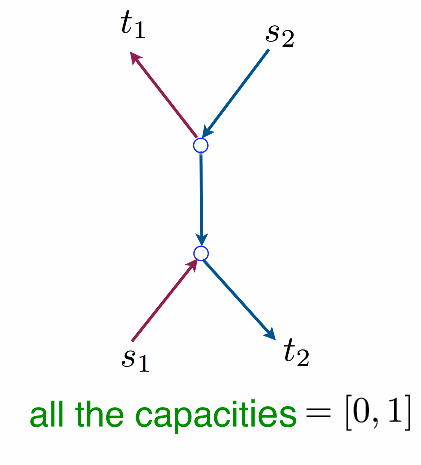} 
   \caption{a simple graph}
\end{figure}

\begin{figure}[htbp]
   \centering
   \includegraphics[bb= 0 0 450 480, scale=0.3]{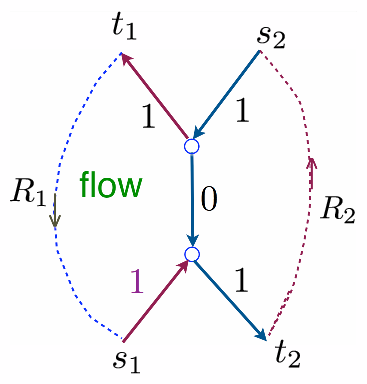} 
   \caption{non-operational cycle with $R_1=R_2=1$}
\end{figure}

\section*{Acknowledgments}
@ The author is very grateful to Prof. Shin'ichi Oishi for  
providing him with  pleasant research facilities during this work.   
 Thanks are   also due to Dinkar Vasudevan for bringing reference
 \cite{korada-vasu} to the author's attention. 


%



%
%


\begin{thebibliography}{999}



\bibitem{al-yeung} R. Ahlswede, N.Cai, S.Y. R. Li and R.W.Yeung, ``Network information flow," {\em IEEE Transactions on Information Theory}, vol.IT-46, no.4, pp. 1204-1216, 2000

\bibitem{yeng-first} R.W. Yeung, {\em A First Course in Information Theory}, Kluwer, 2002


\bibitem{han-cover} T.S. Han,  ``Slepian-Wolf-Cover theorem for a network of channels,"  {\em  Information and Control}, vol.47, no.1, pp. 67-83, 1980

\bibitem{han-book} T.S. Han,   {\em  Information-Spectrum Methods in Information Theory}, 
Springer-verlag, Berlin, 2003


\bibitem{han-shannon} T.S. Han,  ``Multicasting correlated multi-source to multi-sink over a network,"  arXiv:0901.0608, submitted to@{\em IEEE Transactions on Information Theory}


\bibitem{cover} T. M. Cover, ``A simple proof of the data compression theorem of Slepian and Wolf for ergodic sources," {\em IEEE Transactions on Information Theory}, vol.IT-21, pp. 226-228, 1975

\bibitem{verdu-han} S. Verd\'u and T.S.Han, ``A general formula 
	for channel capacity," {\em IEEE Transactions on Information Theory}, 
	vol.IT-40, no.4, pp.1147-1157, 1994

\bibitem{gall} R. G. Gallager, {\em Information Theory and Reliable Communication}, Wiley, New York,1968

\bibitem{cover-thomas} T. M. Cover and J. A. Thomas, {\em Elements of 
Information Theory},  Wiley, New York, 1991

\bibitem{bar-serv} J. Barros and S. D. Servetto, ``Network information flow with correlated sources,"  
{\em IEEE Transactions on Information Theory}, vol.IT-52, no.1,
 pp.155-170, 2006

 
 \bibitem{song} L. Song, R.W.Yeung and N.Cai, ``A separation theorem for single-source network coding,"  
{\em IEEE Transactions on Information Theory}, vol.IT-52, no.5,
 pp.1861-1871, 2006

\bibitem{yan} X. Yan, R.W.Yeung and Z. Zhang, ``The capacity region for multiple-source multiple-sink network coding,"  
{\em Proc. IEEE  International Symposium on Information Theory}, June 2007

\bibitem{song-yeung} L. Song, R.W. Yeung and N.Cai, ``Zero-error network coding for acyclic networks,"  
{\em IEEE Transactions on Information Theory}, vol.IT-49, no.12,
 pp.3129-3139, 2003

 \bibitem{koetter} T. Ho, M. M\'edard, M.Effros and R. Koetter, ``Network coding for correlated sources,"  
{\em Proc.  Conference on Information Science and Systems}, 2004
%
 \bibitem{koetter-1} T. Ho, M. M\'edard, and R. Koetter, D.R.Karger, M.Effros, Jun Shi, and Ben Leong,  ``A random linear network coding approach to multicast,"  
{\em IEEE Transactions on Information Theory}, vol.IT-52, no.10,
 pp.4413-4430, 2006
 
  
 \bibitem{effros} A. Ramamoorthy, K. Jain, A. Chou and M.Effros, ``Separating distributed source coding from network coding,"  
{\em IEEE Transactions on Information Theory}, vol.IT-52, no.6,
 pp.2785-2795, 2006
 
 \bibitem{koetter-med}  R. Koetter and M. Med\'ard,  ``An algebraic approach to network coding,"  
{\em IEEE Transactions on Information Theory}, vol.IT-49, no.5, pp.782-795, 2003

\bibitem{li-yeung} A. Li and R.W. Yeung,
``Network information flow - multiple source," {\em  IEEE International Symposium on Information Theory}, 
2001

%
 \bibitem{xzhang} X. Zhang, Jun Chen, S. B. Wicker and T. Berger, ``Successive coding in multiuser information theory,"  
{\em IEEE Transactions on Information Theory}, vol.IT-53, no.6,
 pp.2246-2254, 2007

 
 \bibitem{imre} I. Csisz\'ar, ``Linear codes for sources and source networks: error exponents, universal coding,"  
{\em IEEE Transactions on Information Theory}, vol.IT-28, no.4,
 pp.585-592, 1982
 
  \bibitem{korn-mart} J. K\"orner and Marton, ``How to encode the modulo-two sum of binary sources," {\em IEEE Transactions on Information Theory}, vol.IT-25, pp. 219-221, 1979



 \bibitem{csis-kor} I. Csisz\'ar and J. K\"orner, {\em Information Theory: Coding Theorems for Discrete Memoryless Systems},  Academic Press, New York, 1981


\bibitem{zhang} X. Yan, J. Yang and Z. Zhang, ``An outer bound for multiple source multiple sink 
coding with minimum cost consideration," {\em IEEE Transactions on Information Theory}, vol.IT-52, no.6, pp. 2373-2385, 2006



\bibitem{meggido} N. Meggido, ``Optimal flows in networks with multiple  sources and multiple  sinks,"  
{\em Mathematical Programming}, vol.7,  pp.97-107, 1974

%
\bibitem{rat-kramer} N. Ratnakar and  G. Kramer, ``The multicast capacity of deterministic relay
networks with no interference,h IEEE Trans. Inform. Theory, vol. 52, no.
6, pp.2425-2432, Jun. 2006.

%
\bibitem{korada-vasu} S. B. Korada and D. Vasudevan,
``Broadcast and Slepian-Wolf multicast over Aref
networks," {\em Proc. IEEE International Symposium on Information Theory}, 
pp. 1656-1660, Toronto, Canada, July 6-11, 2008


\bibitem{avest} S. Avestimehr {\em Wireless network information flow}, 
PhD thesis, University of California, Berkeley, 2008


\bibitem{rama} A. Ramamoorthy,
``Minimum cost distributed source coding over a network," {\em Proc. IEEE International Symposium on Information Theory}, 
Nice, France, 2007; accepted by IEEE Trans. Inform. Theory;
http://arxiv.org/abs/0704.2808

\bibitem{lee} A. Lee, M.Med\'ard, K.Z. Haigh, S. Gowan, and P. Rupel,
``Minimum-cost subgraphs for joint distributed source and network coding," {\em Thirs Workshop on Network Coding, Theory, and Applications}, 
January 2007

\bibitem{avest} S. Avestimehr,
``Minimum cost distributed source coding over a network," {\em Wireless network information flow: adeterministic approach}, PhD thesis, University of California, Berkeley, 2008;
http://www.eecs.berkeley.edu/Pubs/TechRpts/2008/EECS-2008-128.html


%
\end{thebibliography}
\end{document}